\documentclass[a4paper]{PoS}

\title{First numerical experiences with overlap fermions based on the Brillouin kernel}

\ShortTitle{Overlap fermions based on the Brillouin kernel}

\author{\speaker{Stephan D\"urr}\\ 
        University of Wuppertal and IAS/JSC Forschungszentrum J\"ulich, Germany\\
        E-mail: \email{durr\,(AT)\,itp$\,.\,$unibe$\,.\,$ch}}

\author{Giannis Koutsou\\
        Cyprus Institute, CaSToRC, Nicosia, Cyprus\\
        E-mail: \email{g.koutsou\,(AT)\,cyi$\,.\,$ac$\,.\,$cy}}

\abstract{Numerical experiences are reported with overlap fermions which
employ the Brillouin action as a kernel. After discussing the dispersion
relations of both the kernel and the resulting chiral action, some of the
physics features are addressed on quenched backgrounds. We find that the
overlap with Brillouin kernel is much better localized than the overlap with
Wilson kernel. Also a preliminary account is given of the cost of the
formulation, in terms of CPU time and memory.}

\FullConference{34th International Symposium on Lattice Field Theory\\
                24-30 July 2016\\
                University of Southampton, UK}

\renewcommand{\dag}{^\dagger}

\newcommand{\gaf}{\gamma_5}
\newcommand{\nab}{\nabla}
\newcommand{\lap}{\triangle}

\newcommand{\al}{\alpha}
\newcommand{\be}{\beta}
\newcommand{\ga}{\gamma}
\newcommand{\de}{\delta}

\newcommand{\ze}{\zeta}
\newcommand{\et}{\eta}

\newcommand{\rh}{\rho}

\newcommand{\si}{\sigma}

\newcommand{\ps}{\psi}

\newcommand{\bdm}{\begin{displaymath}}
\newcommand{\edm}{\end{displaymath}}
\newcommand{\bea}{\begin{eqnarray}}
\newcommand{\eea}{\end{eqnarray}}
\newcommand{\beq}{\begin{equation}}
\newcommand{\eeq}{\end{equation}}

\newcommand{\mr}{\mathrm}
\newcommand{\mb}{\mathbf}

\begin{document}


\section{Brillouin Fermions}

Designing a lattice fermion action with a continuum-like dispersion relation
along with good chiral properties for $am\ll1$ and small cut-off effects for
$am=O(1)$ remains a challenge.
The ``perfect action'' approach by Peter Hasenfratz and collaborators aimed
at this combination of desirable properties~\cite{Hasenfratz}, and there
have been similar attempts since~\cite{Gattringer,Bietenholz}.
Also the Brillouin action~\cite{Durr:2010ch}
\beq
D_\mr{B}(x,y)=\sum_\mu \ga_\mu \nab_\mu^\mr{iso}(x,y)
-\frac{a}{2}\lap^\mr{bri}(x,y)+m_0\de_{x,y}
-\frac{c_\mr{SW}}{2}\sum_{\mu<\nu}\si_{\mu\nu}F_{\mu\nu}\de_{x,y}
\;,
\label{def_bril}
\eeq
where $\nab_\mu^\mr{iso}$ is a 54-point discretization of the covariant
derivative and $\lap^\mr{bri}$ is a 81-point discretization of the covariant
laplacian, belongs to the same category (it differs from previous attempts in
all coefficients in the stencils being untuned rational numbers).
In order to maintain $\gaf$-her\-miticity an average over all $n$-hop paths
($n=2,3,4$) in the stencil must be taken (cf.\ Sec.\,3 below).


\begin{figure}[b!]
\includegraphics[width=0.5\textwidth]{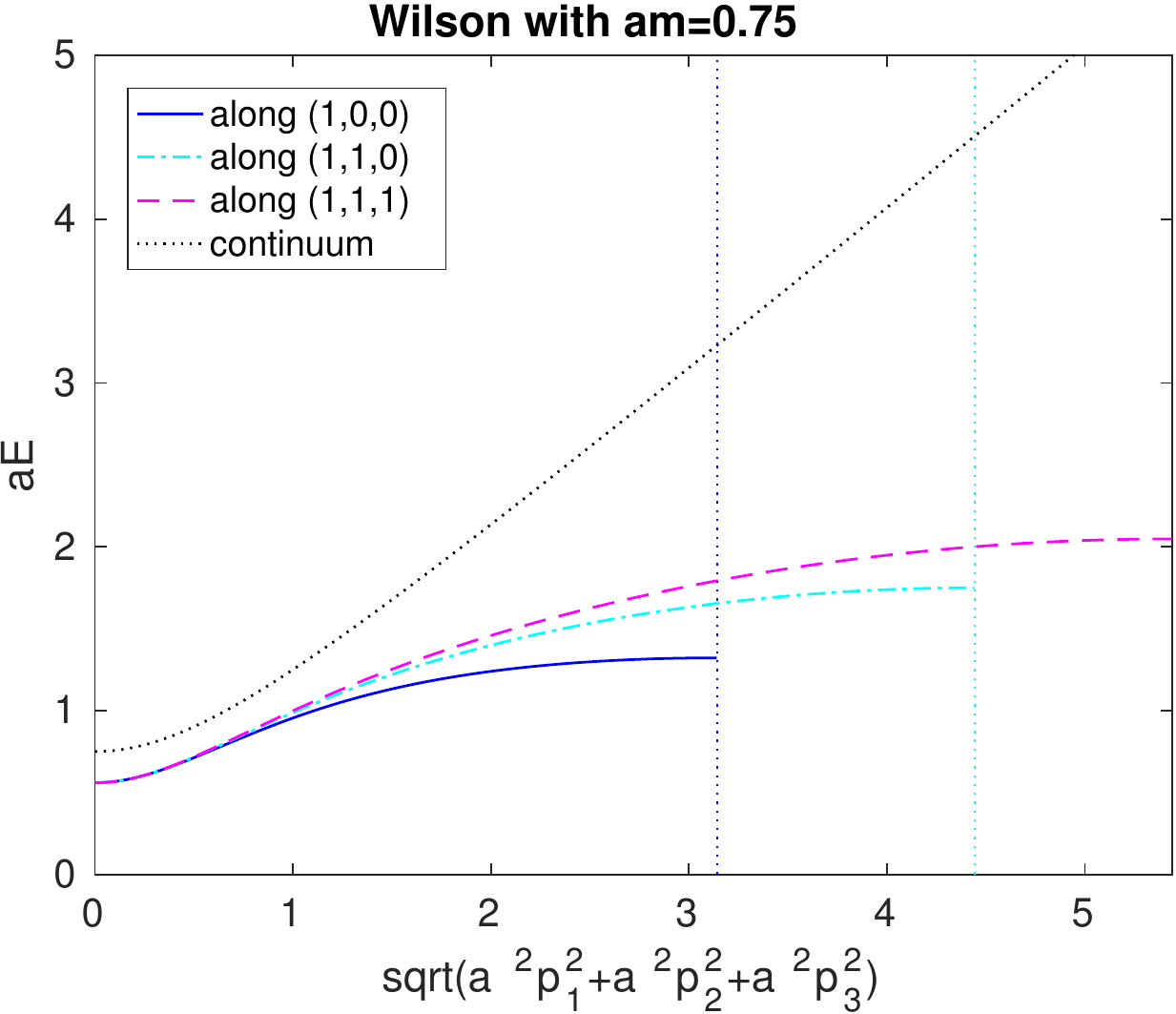}%
\includegraphics[width=0.5\textwidth]{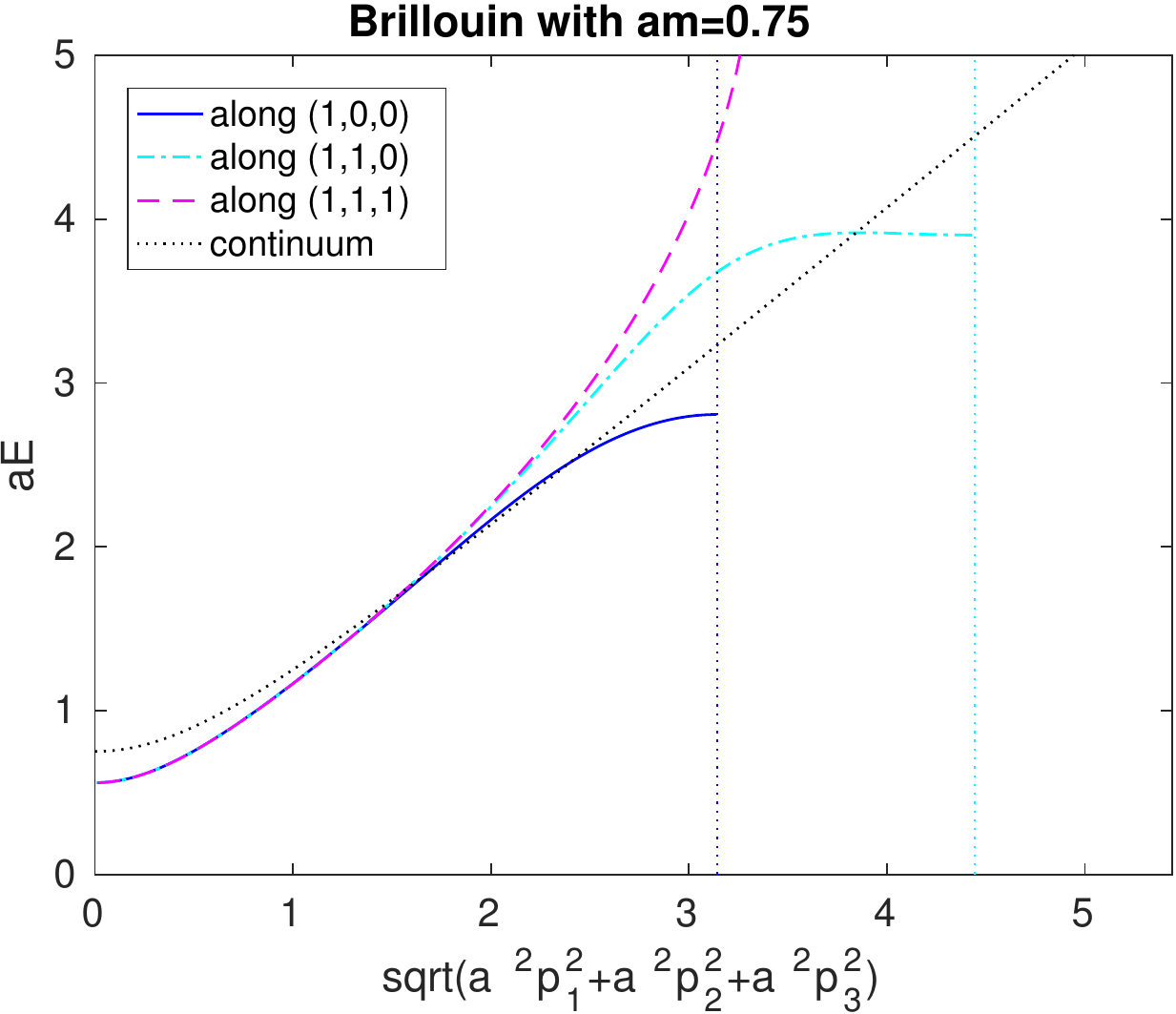}%
\\[2mm]
\includegraphics[width=0.5\textwidth]{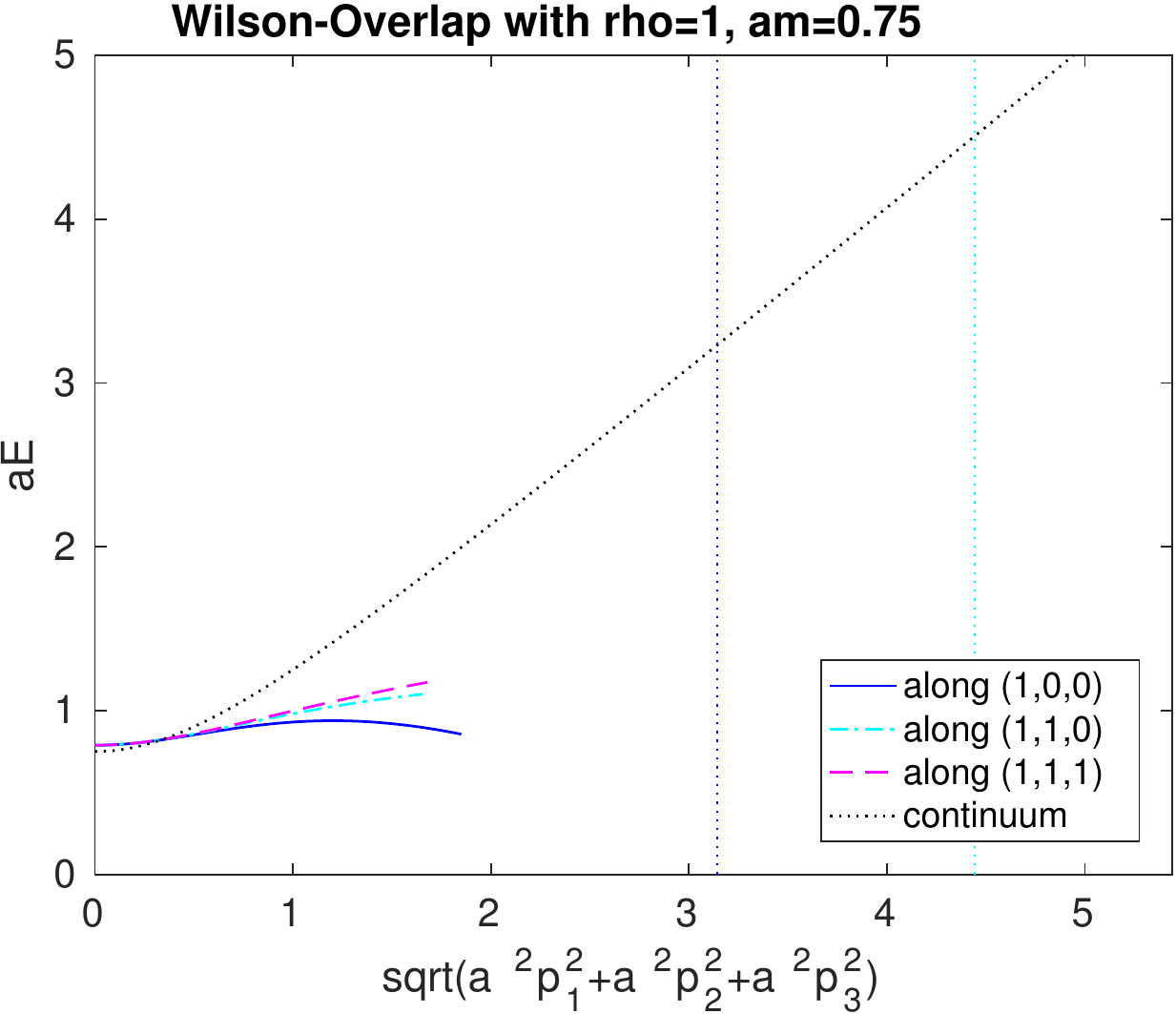}%
\includegraphics[width=0.5\textwidth]{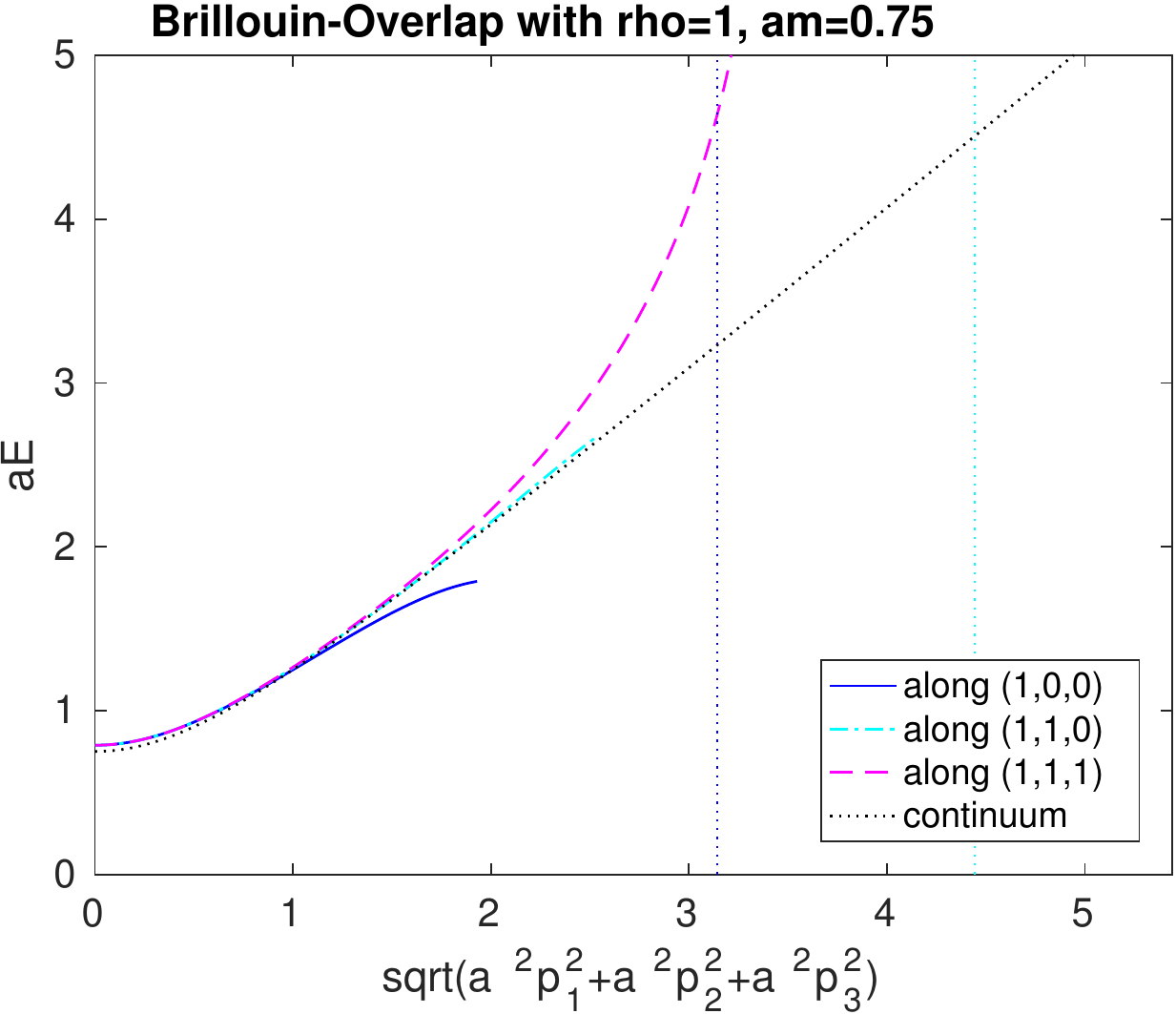}%
\caption{\label{fig:MDR}
Dispersion relations of the Wilson operator (top left), Brillouin operator (top
right), as well as their $\rh=1$ overlap descendents (bottom panels), all at
the rather heavy quark mass $am=0.75$. The Brillouin zone ends at $\pi/a$ on
axis, and it reaches by a factor $\sqrt{2}$ and $\sqrt{3}$ further out for
the other two directions.}
\end{figure}

\section{Quark Dispersion Relations}

The Brillouin action at $am\ll1$ is known to show a good dispersion relation,
both on the quark level (in the weak field limit)~\cite{Cho:2015ffa}, and for
mesons and baryons (in the interacting theory)~\cite{Durr:2012dw}.
Unfortunately, Fig.\,\ref{fig:MDR} shows that this does not carry over to
substantial quark masses; the cut-off effect at $\mb{p}=0$ is as large as with
the Wilson action~\cite{Cho:2015ffa}.
Fortunately, applying the overlap procedure is found to (almost) cure this
deficiency; the massive dispersion relation stays nearly perfect out to
$a||\mb{p}||\simeq2$ with the Brillouin kernel, while it breaks down much
earlier for the Wilson kernel.


\section{Implementation Details of Kernel Action}

It is important to have an efficient implementation of the Brillouin kernel to
start any serious investigation of the Brillouin overlap action.
To this end it seems crucial to precompute all off-diagonal links (2
contributions to 2-hop, 6 contributions to 3-hop, 24 contributions to 4-hop,
always averaged but not necessarily backprojected) and to assemble the relevant
40 links in $W_\nu(x)$ with $\nu=1...40$ to be computed from the
smeared gauge field $V_\mu(x)$ with $\mu=1...4$.

With the object $W_\nu$ in hand the Brillouin flop-count for a matrix-times-vector
operation yields 30192 per site, to be compared with the Wilson flop count of
1368 per site (with mass term in both cases but no gauge compression and no
e/o-decomposition for Wilson)~\cite{slides}.
Hence, the Brillouin-to-Wilson ratio of flops is 22.1, not far from the
measured timing ratio 18.5 on a standard 4-core CPU.
Similarly, the required memory traffic is 3408 floats for Brillouin and 384 for
Wilson, i.e.\ the Brillouin-to-Wilson ratio of traffic is 8.9 (with several rhs
both numbers decrease, but the ratio stays at 8.9)~\cite{slides}.
Overall, the 0.45 bytes/flop sp-ratio of the Brillouin action (as opposed to
1.12 bytes/flop in sp for Wilson) makes it an interesting choice for forthcoming
architectures.
Further details in F2008 are found in the slides~\cite{slides}, and a complete
implementation in C is available at~\cite{github}.

The measured mat-vec timing ratio $\sim\!20$ is mitigated by a factor 4
\cite{forthcoming} (due to reduced iteration count and sub-dominance of scalar
products) in a solver, so the actual cost increase is a factor $\sim\!5$.


\section{Implementation Details of Overlap Action}

\begin{figure}[!b]
\includegraphics[width=0.33\textwidth]{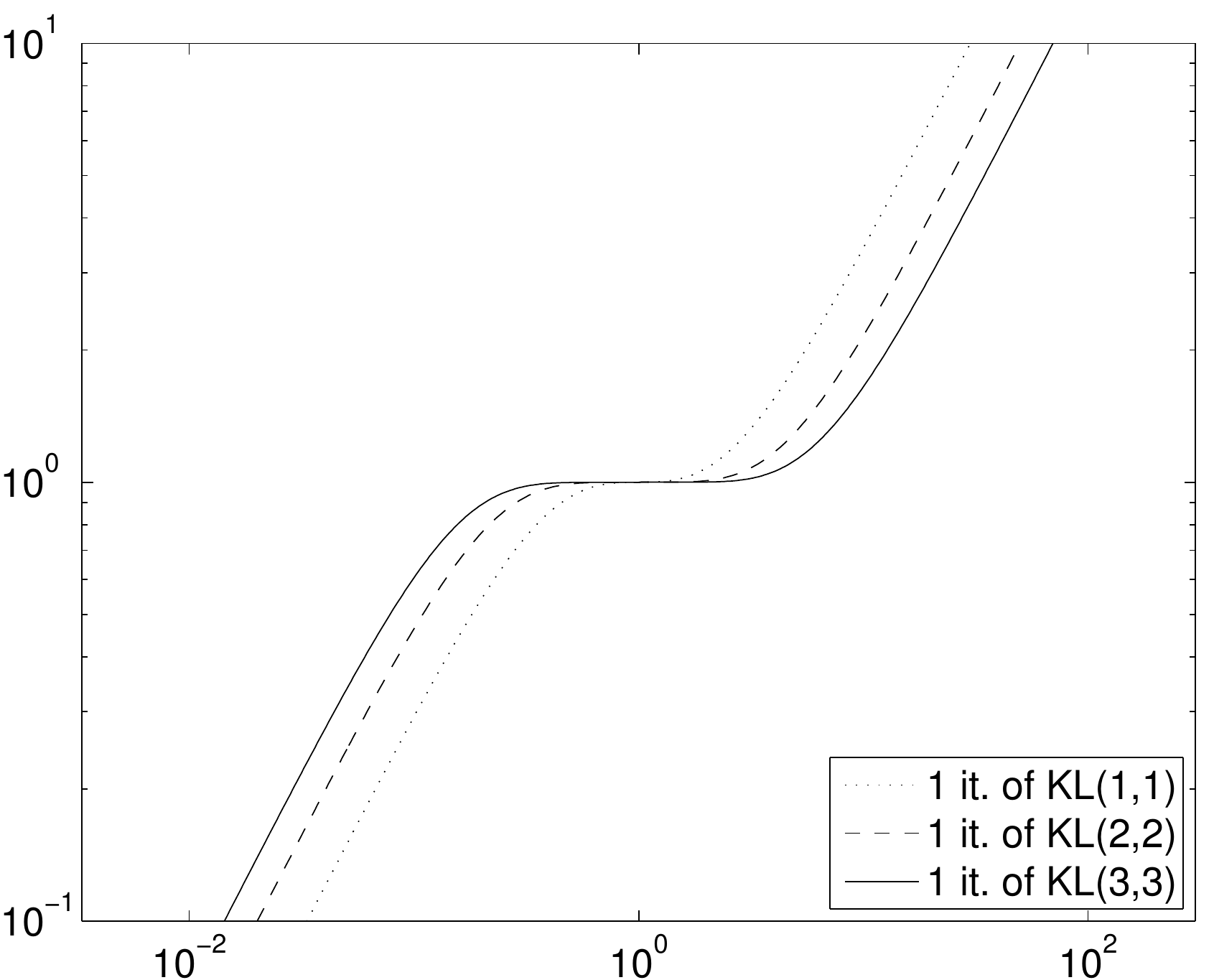}%
\includegraphics[width=0.33\textwidth]{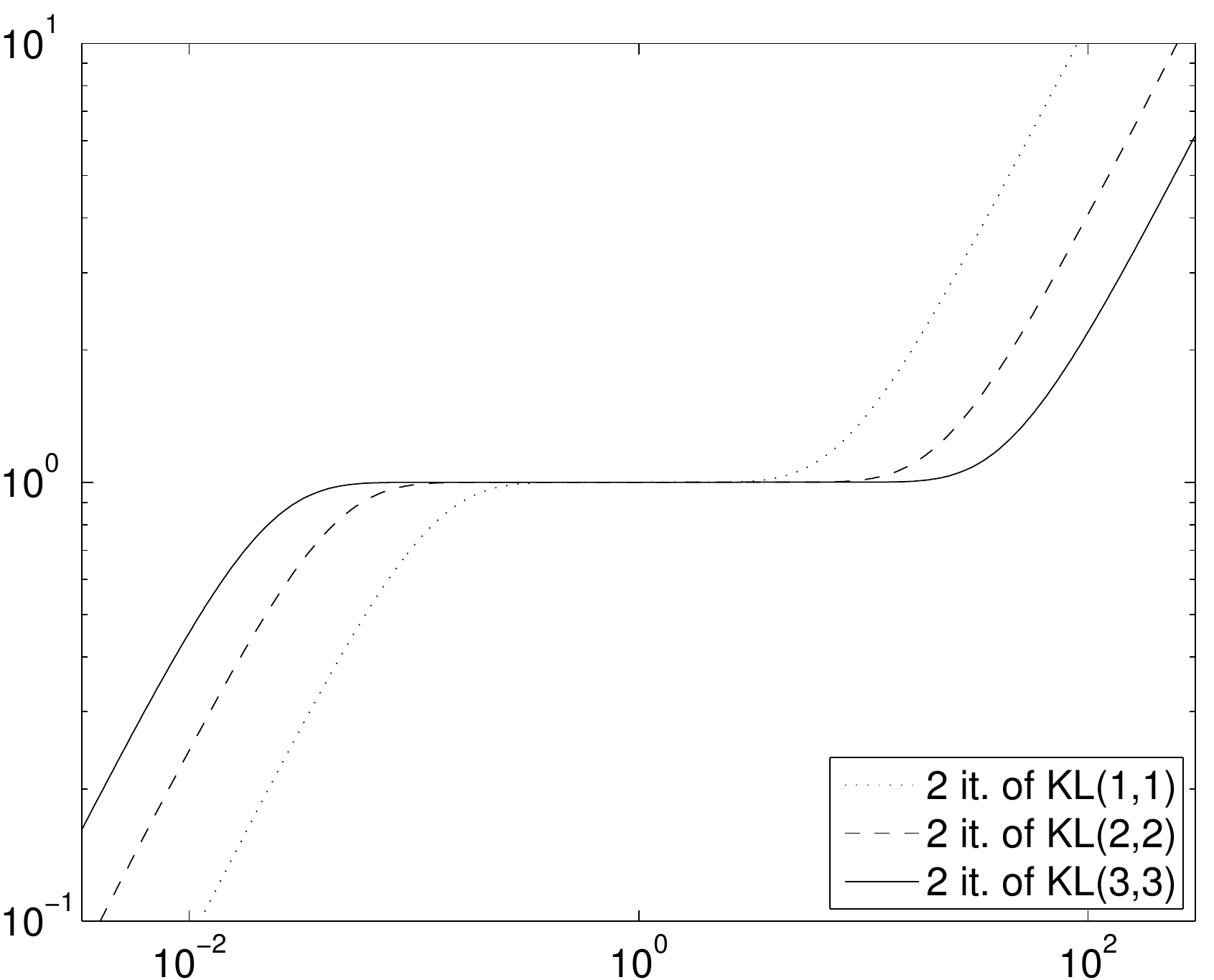}%
\includegraphics[width=0.33\textwidth]{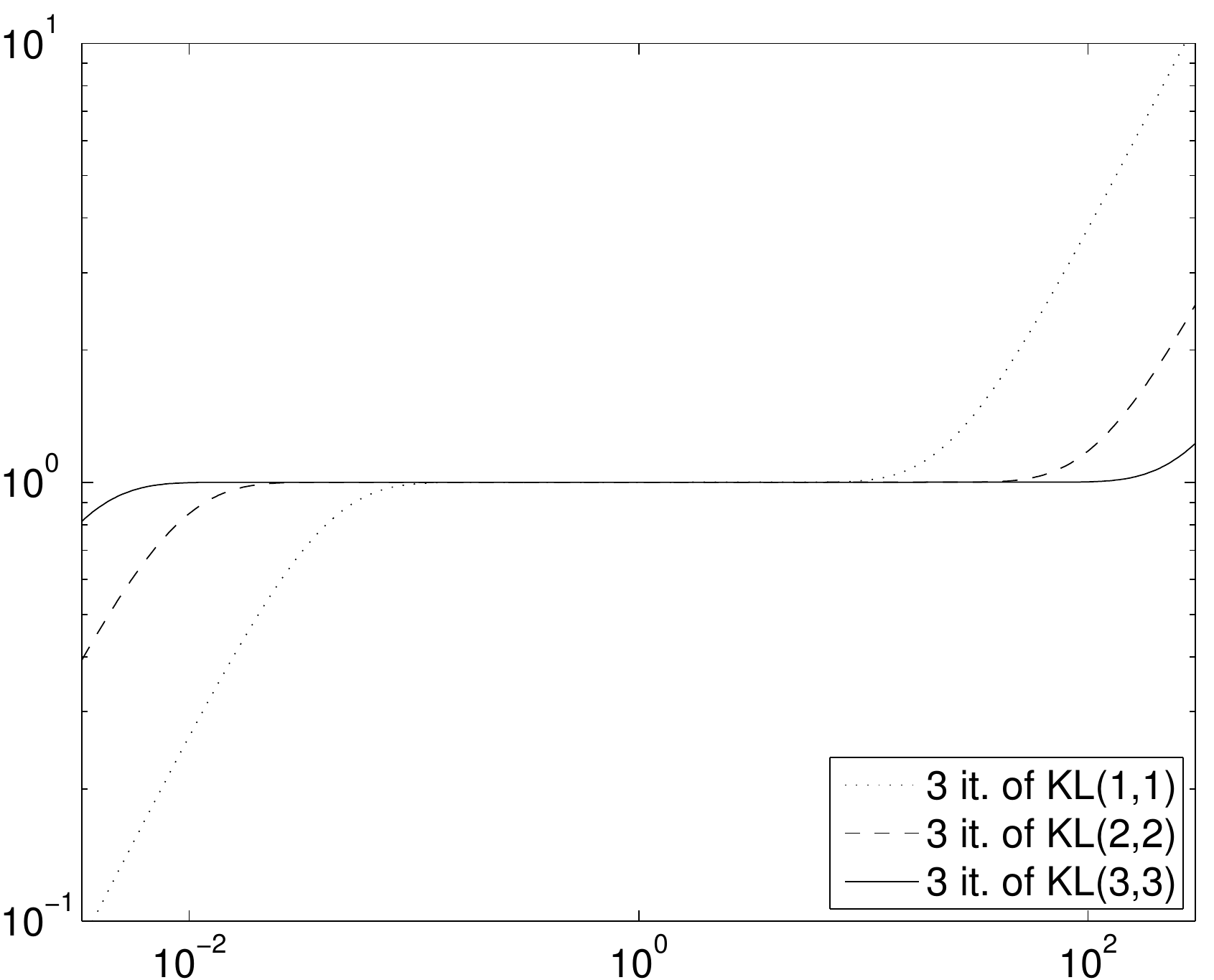}%
\caption{\label{fig:KL}
Approximate Kenney-Laub sign functions $f_{11}, f_{22}, f_{33}$ on $]0,\infty[$
after one (left), two (middle) and three (right) iterations. Note the symmetry
about (1,1), and the strictly monotonic behavior.}
\end{figure}

Starting from a kernel action at negative mass $-\rh/a$ the massless overlap
action is~\cite{Neuberger}
\beq
D_\mr{ovr,0}=\textstyle
\frac{\rh}{a}\big\{1+\gaf\mr{sign}(\gaf D_{\mr{ker},-\rh/a})\big\}
=
\frac{\rh}{a}\big\{1+D_{\mr{ker},-\rh/a}(D_{\mr{ker},-\rh/a}\dag D_{\mr{ker},-\rh/a}^{})^{-1/2}\big\}
\eeq
and the Kenney-Laub family of approximants to the matrix sign function or
inverse square root
\beq
\textstyle
f_{11}(x)=x\frac{3+x^2}{1+3x^2}\;,\;
f_{22}(x)=x\frac{5+10x^2+x^4}{1+10x^2+5x^4}\;,\;
f_{33}(x)=x\frac{7+35x^2+21x^4+x^6}{1+21x^2+35x^4+7x^6}\;,\;
f_{44}(x)=f_{11}(f_{11}(x))
\label{KL}
\eeq
is a convenient but non-optimal choice with remarkable properties (see
Fig.\,\ref{fig:KL} and~\cite{forthcoming}).
This choice does not require any knowledge of the spectral properties of
$\gaf D_{\mr{ker},-\rh/a}$ or $D_{\mr{ker},-\rh/a}\dag D_{\mr{ker},-\rh/a}^{}$,
but if low-lying eigenvalue-eigenvector information is available, it can be
used to speed up the CG used in the partial-fraction expansion of an element in
eq.\,\ref{KL} (see~\cite{forthcoming} for more details).


\section{Overlap Action Properties}

\begin{figure}[b!]
\includegraphics[width=0.5\textwidth]{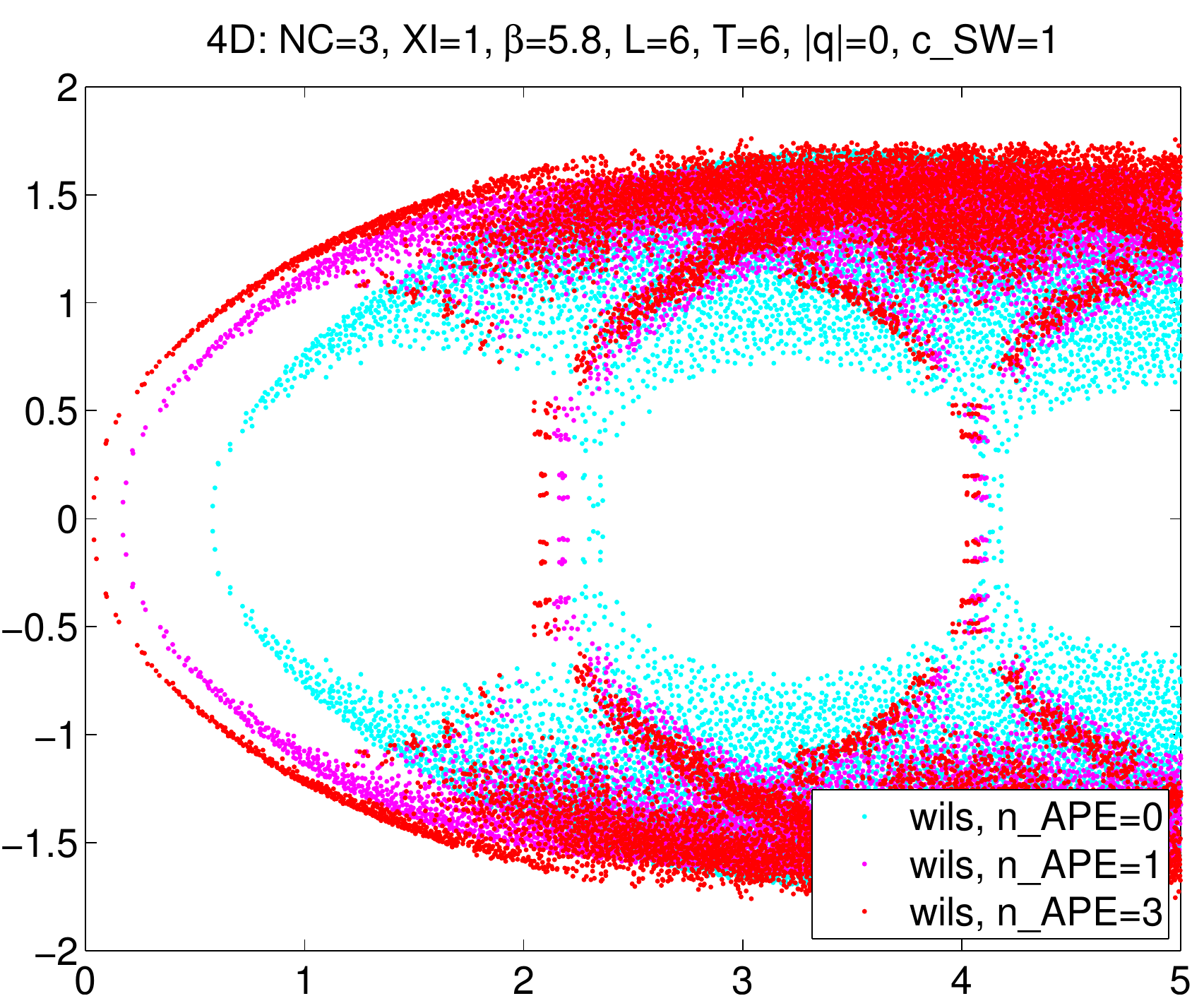}%
\includegraphics[width=0.5\textwidth]{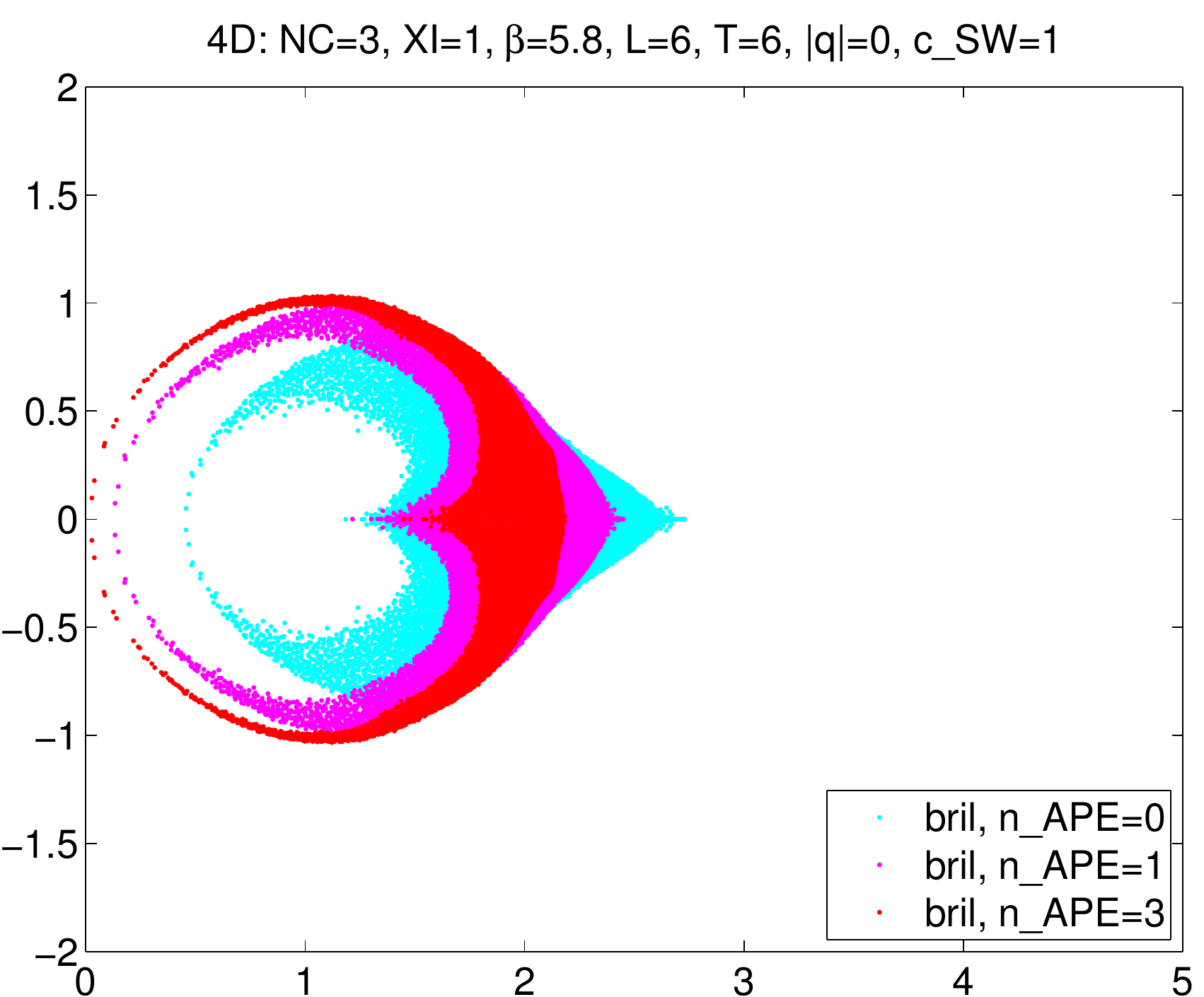}\\
\includegraphics[width=0.5\textwidth]{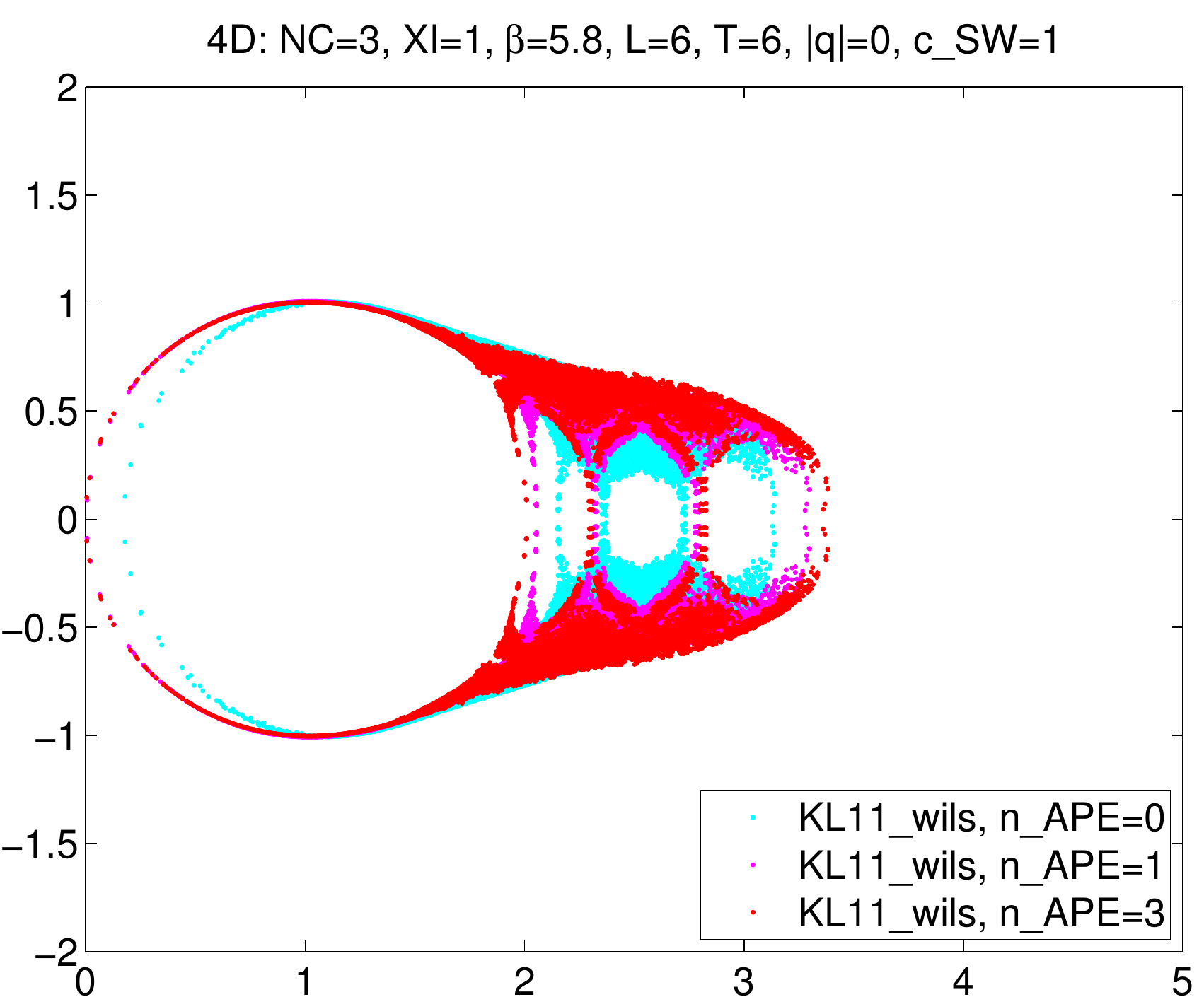}%
\includegraphics[width=0.5\textwidth]{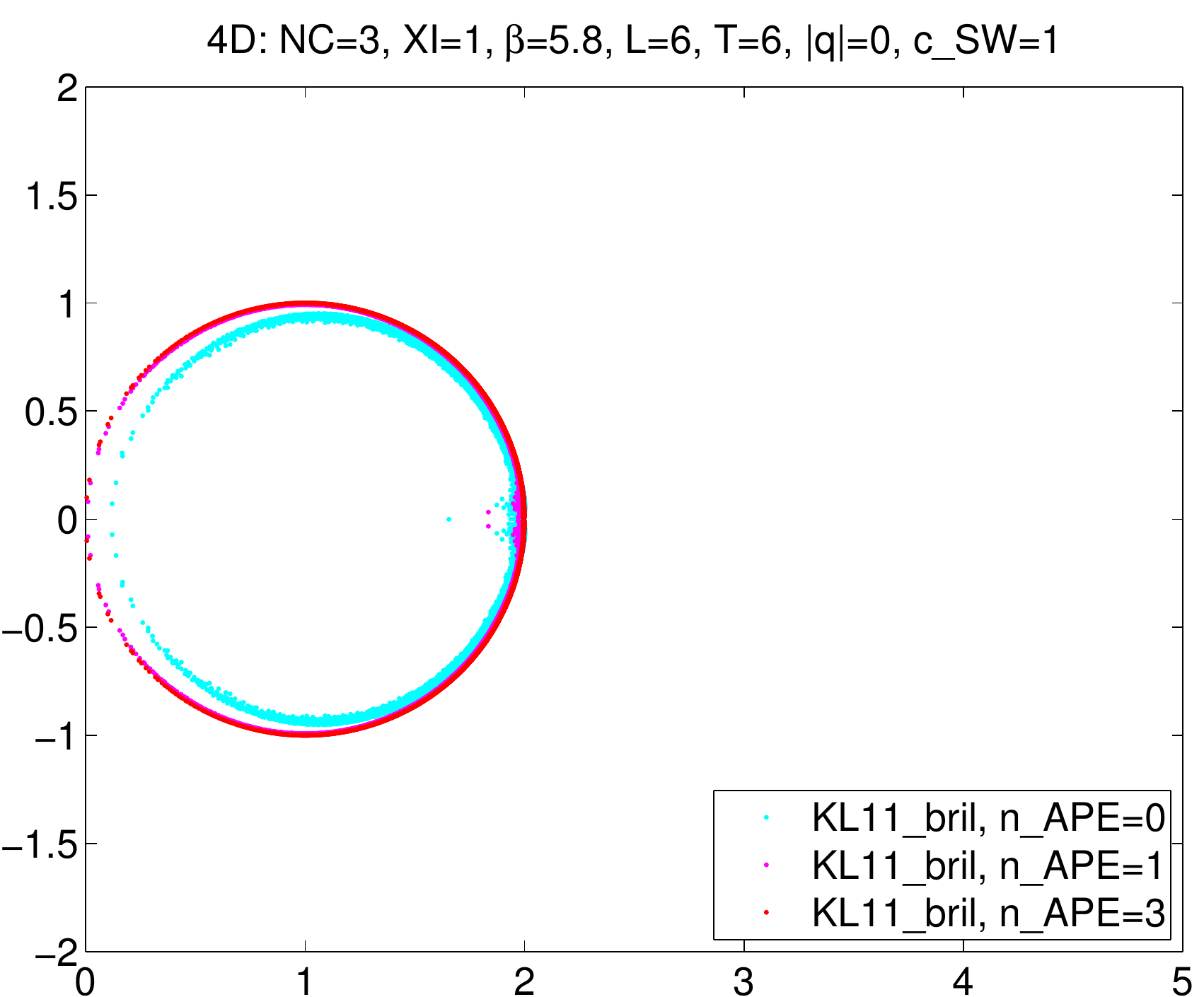}%
\caption{\label{fig:eig}
Eigenvalue spectra of the Wilson (top left) and Brillouin (top right) operators
on a thermalized quenched $6^4$ background, both operators with $c_\mr{SW}=1$.
The three colors refer to three levels of link smearing (cyan for no smearing,
magenta for 1 APE step, red for 3 APE steps, in both cases with $\al=0.72$).
The bottom panels display the eigenvalue spectra of the respective $f_{11}$
overlap approximants ($\rh=1$).}
\end{figure}

In Fig.\,\ref{fig:eig} we show the eigenvalue spectra of the Wilson and
Brillouin actions, and how they are modified if the fixed-order Kenney-Laub
procedure $f_{11}$ from eq.\,(\ref{KL}) is applied on the operator.
With the Wilson kernel the physical branch is being pushed towards
$\mr{Re}(z)=0$ more efficiently than the remaining 15 branches towards
$\mr{Re}(z)=2$.
With the Brillouin kernel even a single iteration of $f_{11}$ seems to
establish an operator with good chiral properties (at least if starting from
$D_\mr{ker}=D_B$ with sufficient link smearing).
Moreover, since the low-lying physical eigenvalues hardly change, it seems like
a self-suggesting idea to use $D_\mr{B}$ or $D_\mr{W}$ to precondition the
$f_{11}$ approximant~\cite{Brannick:2014vda}, and the latter action to
precondition a higher-order approximant, e.g.\ KL44 given by
$f_{44}=f_{11}(f_{11})$~\cite{forthcoming}.


\section{First Spectroscopy Results}

An overlap operator evaluated with infinite precision is normal and satisfies
the Ginsparg-Wilson (GW) relation~\cite{Ginsparg:1981bj}, which was
re-discovered by Peter Hasenfratz (2nd work of Ref.\,\cite{Hasenfratz}).

\begin{figure}[!b]
\includegraphics[width=0.5\textwidth]{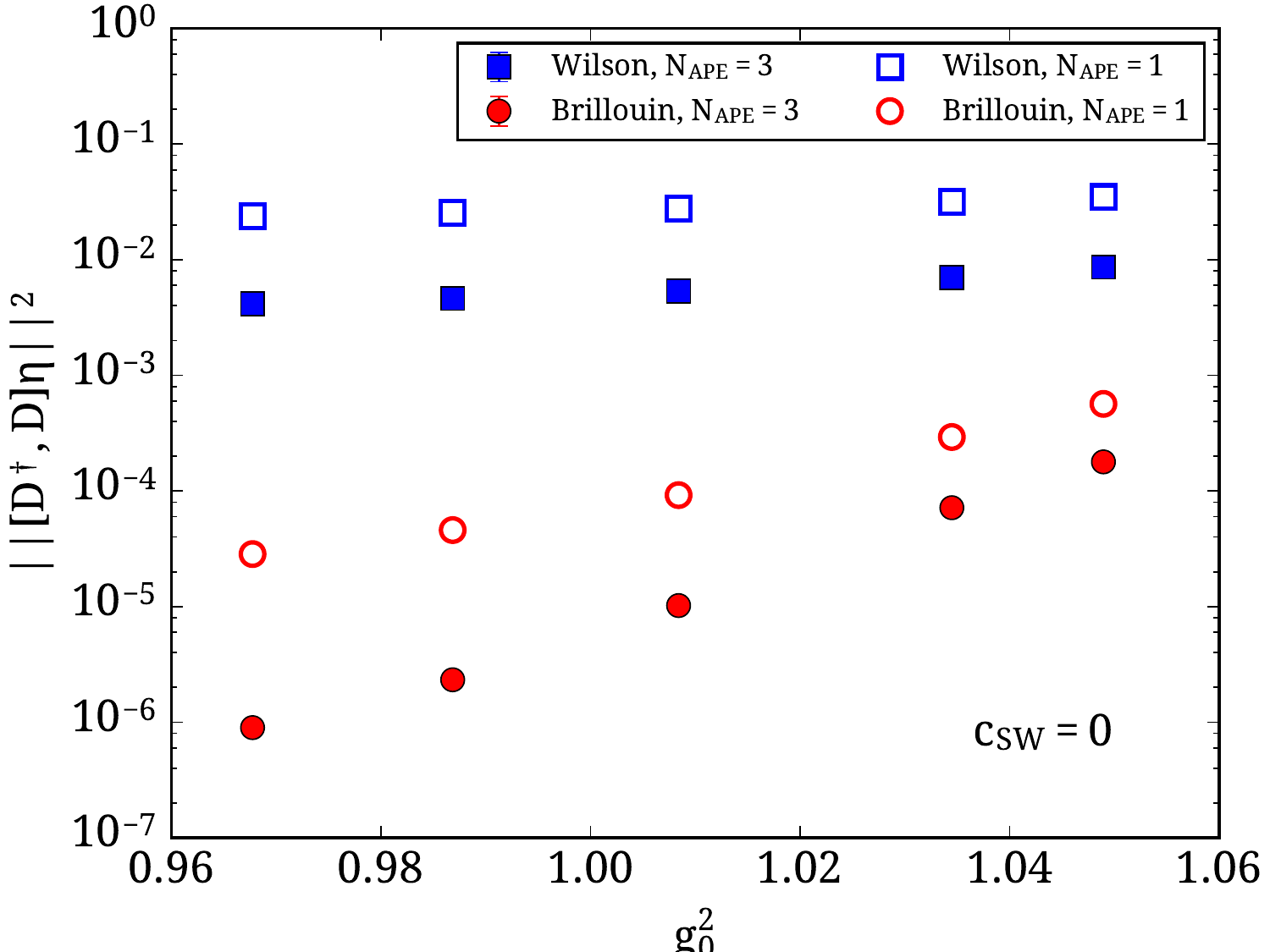}%
\includegraphics[width=0.5\textwidth]{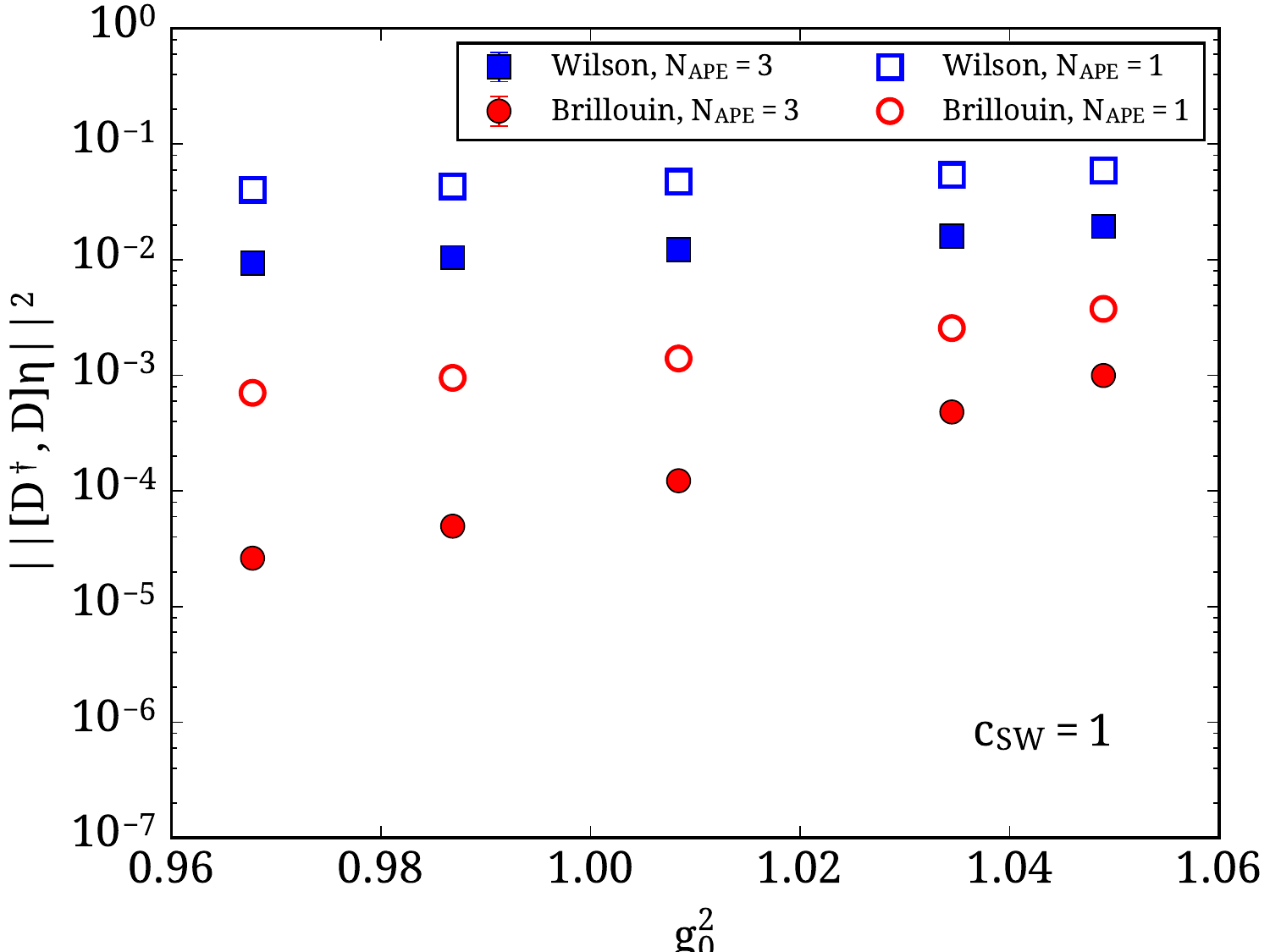}%
\caption{\label{fig:normality}
Remnant non-normality of the KL11 overlap approximants based on the Wilson
kernel (blue squares) and the Brillouin kernel (red circles) versus $6/\be$.
Open symbols refer to 1 step of APE smearing, filled symbols to 3 steps. The
kernel actions at $\rh=1$ may be unimproved (left) or tree-level improved (right).}
\end{figure}

\begin{figure}[!b]
\includegraphics[width=0.5\textwidth]{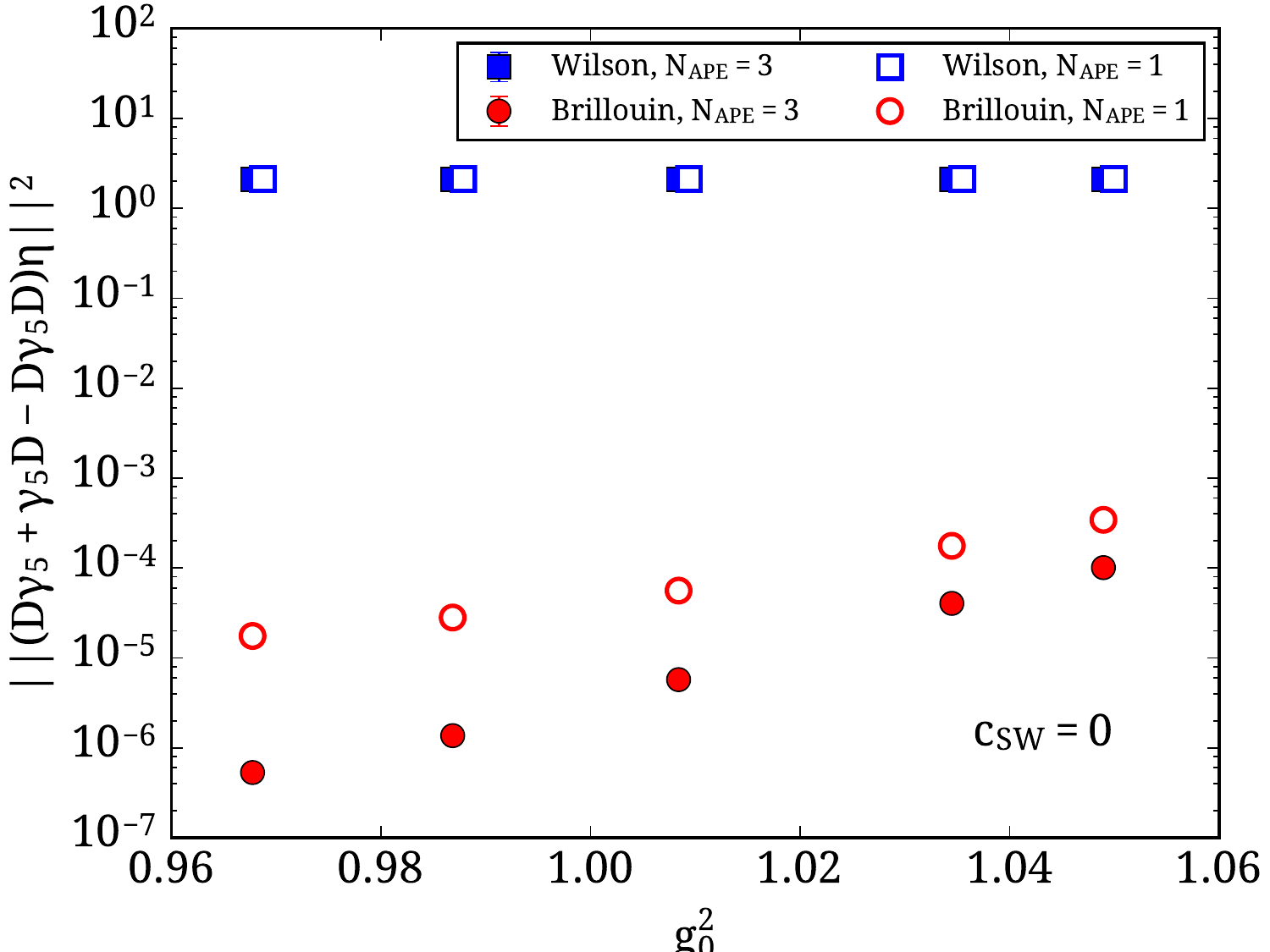}%
\includegraphics[width=0.5\textwidth]{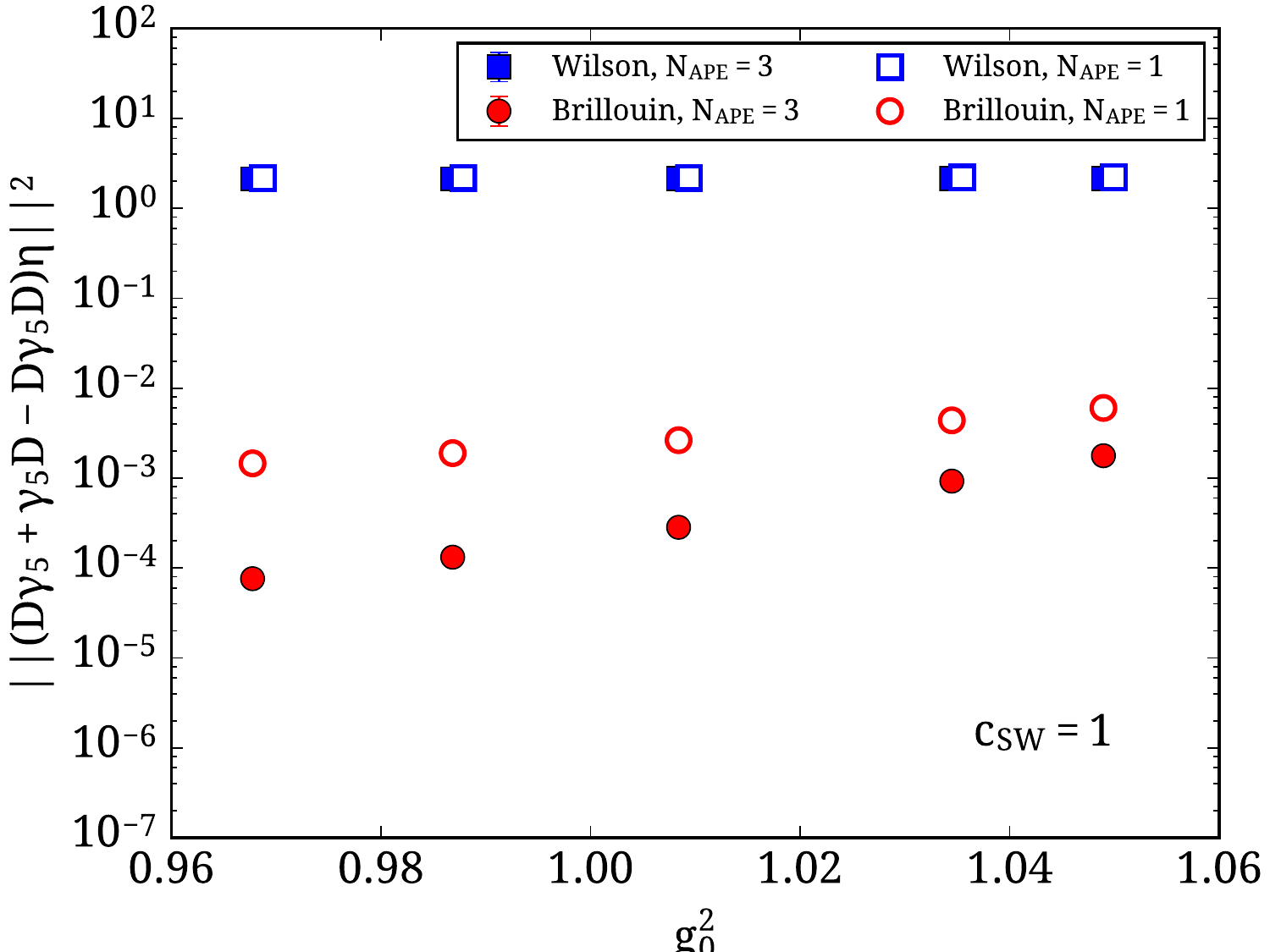}%
\caption{\label{fig:ginspwils}
Remnant violation of the GW-relation of the KL11 overlap approximants based on
the Wilson kernel (blue squares) and the Brillouin kernel (red circles). Open
symbols refer to 1 step of APE smearing, filled symbols to 3 steps. The kernel
actions at $\rh=1$ may be unimproved (left) or tree-level improved (right).}
\end{figure}

In Fig.\,\ref{fig:normality} we plot the remnant non-normality against
$g_0^2=6/\be$ on volume-matched quenched lattices.
As the operator norm is not available for such big matrices, we evaluate for
Gaussian $\et$
\beq
|| (D\dag D - D D\dag) \et ||^2 / ||\et||^2
\eeq
where $D$ denotes the fixed-order KL11 approximant to the massless overlap
action based on either the Wilson (blue) or Brillouin (red) kernel, and $||.||$
denotes the vector 2-norm.
Evidently, the version with the Brillouin kernel fares much better; it also
benefits more from link smearing.

In Fig.\,\ref{fig:ginspwils} we plot the remnant violation of the GW relation
versus $g_0^2=6/\be$ on the same set of lattices.
Again, since the operator norm is not available, we evaluate
\beq
|| (D \gaf + \gaf D - D \gaf D) \et ||^2 / ||\et||^2
\eeq
where $D$ denotes the fixed-order KL11 approximant to the massless overlap
action based on either the Wilson (blue) or Brillouin (red) kernel, and $||.||$
denotes the vector 2-norm.
Once more, the version with the Brillouin kernel performs better and benefits
more from link smearing.

From these figures it seems the Wilson kernels with and without improvement
yield KL11 overlap actions with comparable properties.
On the other hand, among the Brillouin KL11 actions the version with
$c_\mr{SW}=0$ in the kernel seems superior to the version with $c_\mr{SW}=1$
in the kernel.

\begin{figure}[!b]
\hspace*{12mm}%
\includegraphics[width=0.8\textwidth]{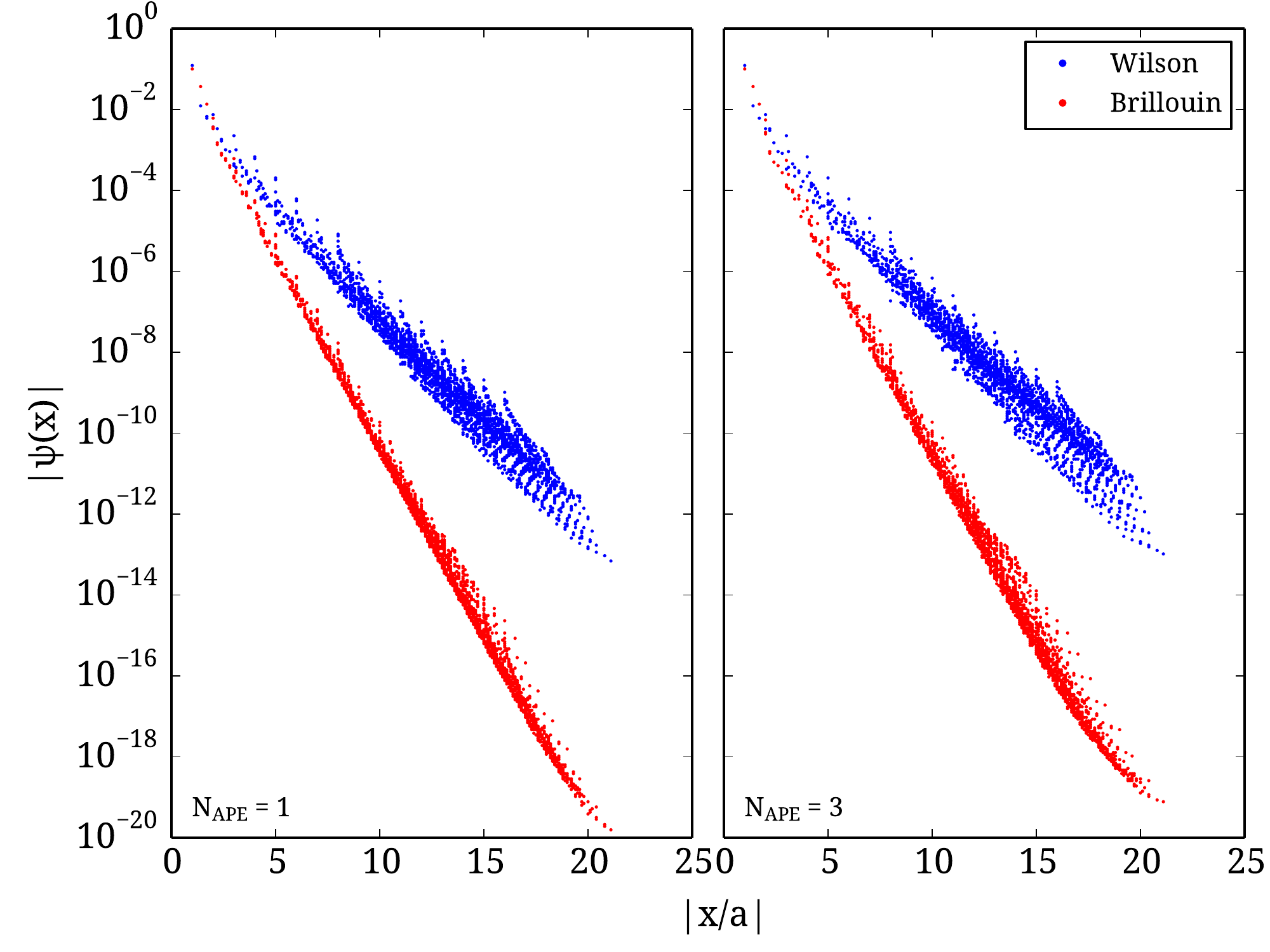}%
\caption{\label{fig:locality}
Fall-off (in coordinate space) of the KL11 approximants to the overlap action
based on the Wilson (blue) and Brillouin (red) kernels, both at $\rh=1$. Using
1 (left) or 3 (right) steps of APE smearing proves immaterial. Note that no
average over different separations $\mb{x}$ with joint $\mb{x}^2$ is taken.}
\end{figure}

The locality properties of the resulting overlap approximants are determined
from a vector $\ze$ which is localized in one grid point, e.g.\ $0$,
but Gaussian in color and spinor space.
We consider the fall-off in the norm of $\ps=D\ze$ which roughly follows
$||\ps(x)||\propto\exp(-\de|x|)$ where $|x|$ is the Euclidean norm in
position space.
Fig.\,\ref{fig:locality} shows that the Brillouin overlap falls about
twice as steep as the Wilson kernel counterpart.
Some meson spectroscopy on $40^3\times64$ lattices is on the slides~\cite{slides}.


\section{Summary and Outlook}

The ``truly perfect action'' (i.e.\ an action which fulfills the physics
requirements of the perfect fermion action as mentioned above
\emph{and} is cheap in terms of CPU requirements) doesn't exist.

The Brillouin kernel improves on the Wilson kernel with respect to
the massless free-field dispersion relation [in the sense that $(aE)^2\simeq
(a\mb{p})^2$ holds over a larger fraction of the Brillouin zone], but it does not bring
any improvement for heavy quark masses [even at $\mb{p}=\mb{0}$].
The overlap procedure (with any kernel) improves both chiral
properties at $am\ll1$ and heavy-quark properties at $am=O(1)$.
The Brillouin overlap action combines the two ingredients and yields a
reasonable approximation to the perfect action as put forth by
P.\,Hasenfratz and other people \cite{Hasenfratz,Gattringer,Bietenholz}.

Our investigation reveals a better normality and reduced GW-violation of a
fixed-order overlap with the Brillouin kernel, in comparison with the Wilson
kernel, and the locality in position space is improved.
In terms of CPU time, the Brillouin matrix-vector multiplication is roughly
a factor 20 more expensive than the Wilson multiplication, but in a solver
about a factor 4 comes back from reduced iteration count and related reasons.
Using a standard cluster architecture, we have been able to invert Brillouin
overlap quarks on lattices of size $40^3\times64$ at reasonable quark masses.

The Brillouin kernel itself shows promising features regarding meson and baryon
dispersion relations \cite{Durr:2012dw}.
We plan to repeat this kind of investigation for the Brillouin-overlap action.

{\bf Acknowledgements}: Numerical computations were performed on JUROPA at JSC.




\begin{thebibliography}{99}

\bibitem{Hasenfratz}
  T.~A.~DeGrand, A.~Hasenfratz, P.~Hasenfratz and F.~Niedermayer,
  Nucl.\ Phys.\ B {\bf 454}, 587 (1995)
  [hep-lat/9506030].
  P.~Hasenfratz,
  Nucl.\ Phys.\ Proc.\ Suppl.\  {\bf 63}, 53 (1998)
  [hep-lat/9709110].
  P.~Hasenfratz,
  Nucl.\ Phys.\ B {\bf 525}, 401 (1998)
  [hep-lat/9802007].

\bibitem{Gattringer}
  C.~Gattringer,
  Phys.\ Rev.\ D {\bf 63}, 114501 (2001)
  [hep-lat/0003005].
  C.~Gattringer, I.~Hip and C.~B.~Lang,
  Nucl.\ Phys.\ B {\bf 597}, 451 (2001)
  [hep-lat/0007042].

\bibitem{Bietenholz}
  W.~Bietenholz and U.~J.~Wiese,
  Nucl.\ Phys.\ B {\bf 464}, 319 (1996)
  [hep-lat/9510026].
  W.~Bietenholz and I.~Hip,
  Nucl.\ Phys.\ B {\bf 570}, 423 (2000)
  [hep-lat/9902019].
  W.~Bietenholz,
  Eur.\ Phys.\ J.\ C {\bf 6}, 537 (1999)
  [hep-lat/9803023].

\bibitem{Durr:2010ch} 
  S.~Durr and G.~Koutsou,
  Phys.\ Rev.\ D {\bf 83}, 114512 (2011)
  [arXiv:1012.3615 [hep-lat]].

\bibitem{Cho:2015ffa} 
  Y.~G.~Cho, S.~Hashimoto, A.~Juttner, T.~Kaneko, M.~Marinkovic, J.~I.~Noaki and J.~T.~Tsang,
  JHEP {\bf 1505}, 072 (2015)
  [arXiv:1504.01630 [hep-lat]].

\bibitem{Durr:2012dw} 
  S.~Durr, G.~Koutsou and T.~Lippert,
  Phys.\ Rev.\ D {\bf 86}, 114514 (2012)
  [arXiv:1208.6270 [hep-lat]].

\bibitem{slides}
See slides at https://conference.ippp.dur.ac.uk/event/470/session/13/contribution/64

\bibitem{github}
See details at https://github.com/g-koutsou/qpb

\bibitem{forthcoming}
S.~Durr and G.~Koutsou, forthcoming.

\bibitem{Neuberger}
  H.~Neuberger,
  Phys.\ Lett.\ B {\bf 417}, 141 (1998)
  [hep-lat/9707022].
  H.~Neuberger,
  Phys.\ Lett.\ B {\bf 427}, 353 (1998)
  [hep-lat/9801031].

\bibitem{Brannick:2014vda} 
  J.~Brannick, A.~Frommer, K.~Kahl, B.~Leder, M.~Rottmann and A.~Strebel,
  Numer.\ Math.\  (2015)
  [arXiv:1410.7170 [hep-lat]].

\bibitem{Ginsparg:1981bj} 
  P.~H.~Ginsparg and K.~G.~Wilson,
  Phys.\ Rev.\ D {\bf 25}, 2649 (1982).

\end{thebibliography}
\end{document}